# Novel primary photoexcitations in π-conjugated donor-acceptor copolymers probed by transient magneto-photoinduced-absorption

Uyen N. V. Huynh[1], Tek P. Basel[1], L. Dou[2], Karan Aryanpour[3], Gang Li[2], Sumit Mazumdar[3], Eitan Ehrenfreund[4], Yang Yang[2], and Z. Valy Vardeny[1*]

[1]Department of Physics & Astronomy, University of Utah, Salt Lake City, Utah 84112

[2]Department of Materials Science & Engineering, University of California-Los Angeles, Los Angeles, California 90095

[3]Department of Physics, University of Arizona, Tucson Arizona 85721

[4]Department of Physics, Technion-Israel Institute of Technology, Haifa 32000, Israel

*Correspondence to:  val@physics.utah.edu

**The saga of the primary photoexcitations in π-conjugated polymers has been a source of extraordinary scientific curiosity that has lasted for more than three decades. From soliton excitations in trans-polyacetylene, to singlet and triplet excitons and polarons in other polymers, to charge transfer excitons in blends of polymers and fullerenes, the field has been rich with a variety of different photoexcitation species. Here we show the photogeneration of a novel primary intrachain photoexcitation species, namely the composite multi-exciton (CME) in π-conjugated donor-acceptor (DA)-copolymers used in organic photovoltaic (OPV) solar cells. We utilized the magnetic field response of the transient photoinduced absorption from sub-picosecond to millisecond to show in pristine DA-copolymer early photogeneration of the CME species that is composed of four coupled spin ½ particles, having unique optical and magnetic signatures. This species decomposes into two independent triplets in the microsecond time domain. Importantly in copolymer/fullerene blends the CME ionization generates photocarriers by a unique process that may enhance the photocurrent in OPV solar cells.**

The field of 'photoexcitations in π-conjugated polymers (PCP)' has been debated since as early as 1980 with the prediction[1] that charge solitons[2] are photogenerated upon photon absorption in the ground state degenerate trans-$(CH)_x$[1,2]. The debate heated up when the nature of the primary photoexcitations, namely free carriers vs. excitons, was considered in non-degenerate PCP such as poly(*para*-phenylene)-vinylene (PPV) and polythiophene (PT)[3-5]. This debate took a new twist when the exciton dissociation in PCP/fullerene blends was discussed[6,7], since this process has bearing on potential applications in OPV solar cells[8]. The field has been further enriched since the demonstration of singlet fission[9], which is a spin allowed process in which a singlet exciton (SE) converts into two independent triplet excitons (TE) via dissociation of an intermolecular state of triplet-triplet (TT-pair)[10-12]. In the present work we discovered a novel primordial photoexcitation species in π-conjugated DA-copolymers, that we dub composite multiple exciton (CME), which is composed of four coupled spin ½ particles having unique optical and magnetic signatures. The CME may dissociate in DA-copolymer/fullerene blend by a unique process that involves electron-hole polaron pair in the triplet spin configuration ($PP_T$), which may increase the photocurrent in OPV solar cells. In our studies we have used, for the first time, the ideal tool for identifying such species, namely, the transient magnetic photoinduced absorption (t-MPA) technique, which is the magnetic field effect (M-) on the transient photo-induced absorption (t-PA) spectrum.

The DA-copolymer chain contains two different organic moieties with different electron affinities (Fig. 1a) that play the role of electron donor (D) and acceptor (A), respectively[13-18]. This intrachain D-A character leads to lower band-gap ($E_g$) than that in more traditional PCPs such as PPV and PT, and therefore can absorb more photons from the solar spectrum[19]. Indeed some DA-copolymer/fullerene blends that show extended absorption to the near-infrared have been shown to generate record high OPV solar cell efficiency (>8%) based on a single active layer device[20]. However, very recently a DA-copolymer with higher $E_g$ has shown even larger OPV efficiency (>10%) when blended with fullerenes[21]. This indicates that the small $E_g$ alone cannot explain the high efficiency of OPV cells based on DA-copolymers, emphasizing that the nature of the primary photoexcitations in these copolymers needs be better understood. This is especially intriguing since the small $E_g$ value in DA-copolymers ( 1.4-1.7 eV) and the rather robust energy difference, $\Delta_{ST}$ (~0.7 eV) known to exist between the SE and TE energies in PCPs[22] form resonance between the lowest SE (at energy $E_S$) and TT-pair (at twice the TE energy ($2E_T$)). This resonance may lead to *strong coupling* between the two states; consequently novel primary photoexcitations may occur[18].

This is the case for the π-conjugated PDTP-DFBT DA-copolymer that we have studied here (see structure in Fig. 1a inset)[17,20]; as well as in two other DA-copolymers[15,16] with different D/A moieties that we have also investigated[18,23]. The PDTP-DFBT backbone structure does not possess the symmetry elements of traditional PCPs[24]. Nevertheless, for the sake of convenience we use here the notations compatible with the PCP irreducible representations to describe the excited states of the DA-copolymer[18]. These include the $1^1B_u$ (the odd parity SE); $m^1A_g$ (the even parity state most strongly coupled to the $1^1B_u$); and their counterparts in the triplet manifold, namely $1^3B_u$ and $m^3A_g$. In addition, the $2^1A_g$ state (that is traditionally considered to be an intrachain TT-pair state), which is forbidden in most PCPs becomes partially allowed in DA-copolymers, because of lacking of inversion symmetry in the backbone chain[18].

The PDTP-DFBT absorption and photoluminescence (PL) spectra are shown in Fig. 1a. The Stokes shifted 0-0 PL band peaks at 1.38 eV, considerably lower than in any traditional PCP[25]. In order to more precisely determine the energies $E(1^1B_u)$ [$=E_g$] and $E(m^1A_g)$ we measured the electro-absorption (EA) spectrum of pristine PDTP-DFBT film deposited on an inter-digitated electrode substrate, subjected to a modulated applied voltage at frequency *f* (see *Supplement information* (S.I.)-S2). In general, the EA spectrum of PCPs shows two dominant optical features; a derivative-like Stark effect feature at $E(1^1B_u)$, and a field-induced absorption at $E(m^1A_g)$ due to the partial symmetry breaking associated with the applied field[24]. The EA spectrum of PDTP-DFBT (Fig. 1b) indeed exhibits these spectral signatures; a derivative-like feature with zero-crossing at ~1.55 eV, which we identify as $E(1^1B_u)$, and a positive band with 0-0 at ~1.95 eV, which we assign as $E(m^1A_g)$ (see Fig. 1d). The energy difference, $\Delta E = E(m^1A_g) - E(1^1B_u)$ has been traditionally used to estimate the exciton binding energy[24]; in PDTP-DFBT we get from the EA spectrum analysis $\Delta E \approx 0.4$ eV, which is considerably lower than that in more traditional PCPs[26]. We note that $\Delta E$ is also expected to be the transition energy of the PA band from the photogenerated $1^1B_u$ into the $m^1A_g$ ($1^1B_u \rightarrow m^1A_g$), namely $PA_{SE}$ (see Fig. 1d)[27].

Figure 1c depicts the steady state PA (ss-PA) spectrum in a film of solid state solution, in which isolated pristine PDTP-DFBT chains are embedded in polystyrene (see S.I.-S1). The PA spectrum was measured at 1 kHz modulation frequency at 300K, using the background PA spectrum in the ps pump-probe measurement (S.I.-S2). The spectrum is dominated by a single PA band ($PA_T$) that peaks at ~0.95 eV, which we assign, as in traditional PCPs, to the strongest transition in the triplet manifold, namely $1^3B_u \rightarrow m^3A_g$ (Fig. 1d)[27]. We performed PL-detected magnetic resonance (PLDMR) (Fig. 1e) and magnetic field dependent PA (ss-MPA; Fig. 1f)) (see S.I.-S3) to identify the spin state of these long-lived photoexcitation species. The PLDMR($B$) response shows a 'full-field' powder pattern around $B_0$=1010 Gauss, which is typical to TE[28]. From the PLDMR powder pattern we can determine the zero-field splitting (ZFS) parameters, $D$ and $E$ of isolated TE in the PDTP-DFBT copolymer chains. In general, spin triplet full-field powder pattern has singularities[28] at $B_0 \pm D$ and peaks at $B_0 \pm (D \pm 3E)/2$, and thus we obtain from the PLDMR($B$) response $D$=38 mT and $E$=15 mT. We also performed steady-state magneto-PA (ss-MPA($B$)), where MPA=[PA($B$)-PA(0)]/PA(0) and $B$ is the magnetic field, at the $PA_T$ band (Fig. 1f), which shows a typical response of triplet excitons[29]. In fact, the ss-MPA($B$) response can be well fit (Fig. 1f) using the same ZFS parameters extracted from the PLDMR($B$) response (see S.I.- S4(ii)).

Since we determined E($m^1A_g$)≈1.95 eV from the EA spectrum, we can estimate its triplet counterpart, E($m^3A_g$) that is lower by about 0.2 eV[27], namely E($m^3A_g$)≈1.75 eV. Consequently, from E($m^3A_g$) and $PA_T$ transition energy in the triplet manifold we can determine the energy of the lowest triplet exciton in PDTP-DFBT, $E_T$=E($1^3B_u$)=E($m^3A_g$)-E($PA_T$)≈1.75-0.95≈0.8 eV (see Fig. 1d). This value is also in agreement with an alternative estimation starting from E($1^1B_u$), since the energy gap, $\Delta_{ST}$=($1^1B_u$)–E($1^3B_u$) in PCP is of the order of 0.7-0.8 eV[22]; thus $E_T$=($1^1B_u$)-$\Delta_{ST}$≈0.8 eV. It is thus clear that the lowest singlet in PDTP-DFBT (=1.55 eV) is nearly resonant with twice the lowest triplet (2x0.8=1.6 eV), i.e. E($1^1B_u$)≈$2E_T$≈E($2^1A_g$) (see Fig. 1d), which calls for strong interaction between the lowest SE and TT states in this copolymer[18]. In addition, the inherent absence of inversion symmetry in the DA-copolymer may in fact enhance this interaction. We show here by employing the t-MPA technique that this interaction leads to a novel primary photoexcitation species that is composed of both SE and TT components, having unique optical and magnetic properties.

We discuss the picosecond (ps) transient spectroscopy of PDTP-DFBT by first examining the mid-IR ps t-PA spectrum of a traditional PCP, which is a soluble derivative of PPV, namely DOO-PPV[27] as presented in Fig. 2a. The ps t-PA measurements were performed using the polarized pump-probe technique with 300 fs time resolution, as described in the *Methods* section. The t-PA spectrum of DOO-PPV contains a single PA band ($PA_{SE}$) due to the SE that peaks at 0.95 eV, which is close to the energy difference, $\Delta E = E(m^1A_g) - E(1^1B_u)$ in this polymer[30]. This PA band is correlated with the stimulated emission band of DOO-PPV and decays with a time constant of ~200 ps, in agreement with the PL quantum efficiency (PLQE) of this polymer (~20%)[30,31].

In contrast, the t-PA spectrum of the PDTP-DFBT copolymer film of solid state solution in polystyrene (Fig. 2b) exhibits *two PA bands*; namely $PA_1$ at 0.4 eV and $PA_2$ at 0.82 eV, respectively, which are formed within our experimental time resolution (~300 fs). The two PA bands decay together with time constant of ~ 30 ps showing the same dynamics (Fig. 2c), which is in agreement with the PLQE of ~3% that we measured in neat films. We also observed transient photoinduced dichroism for the two PA bands (Fig. 2d), namely $-\Delta T_{\parallel} \neq -\Delta T_{\perp}$, where $-\Delta T_{\parallel}$ ($-\Delta T_{\perp}$) is the PA measured for the pump-probe polarizations parallel (perpendicular) to each other. The degree, $P$ of 'linear polarization memory' is defined as $P(t) = (\Delta T_{\parallel} - \Delta T_{\perp})/(\Delta T_{\parallel} + \Delta T_{\perp})$. As seen in Fig. 2d inset, $P$ at $t=0$ [namely $P(0)$] of the two PA bands is not the same; $P(0)=0.33$ for $PA_2$, but is only 0.22 for $PA_1$. However as seen in Fig. 2d, $P(t)$ decays similarly for the two PA bands i.e. within ~50 ps. From the same PA decay dynamics and $P(t)$ kinetics we conclude that the two PA bands originate from the same species that is composed of two correlated components, indicating that this photoexcitation is an unusual composite particle, which we identify as composite multi-exciton (CME). The following t-MPA studies further characterize the CME state.

First we note that $PA_1$ peaks at 0.4 eV, which is $\Delta E$ between $E(1^1B_u)$ and $E(m^1A_g)$ in PDTP-DFBT. We therefore assign it to the transition '$1^1B_u \rightarrow m^1A_g$' (Fig. 1d) as in other, more traditional PCPs; but here this transition is from the lowest CME state that is a correlated state of the SE and TT-pair[18]. We also assign $PA_2$ at 0.82 eV to a strong transition in the TT manifold (TT$\rightarrow$TT* in Fig. 1d) that originates from the same CME lowest state. We show below using the transient magneto-PA (t-MPA) technique (where t-MPA=[t-PA($B$)- t-PA(0)]/t-PA(0), see

*Methods*), that the t-MPA(*B*) responses of the two PA bands are correlated to each other. This indicates that the two PA bands belong to a single photoexcitation species, which is the CME.

Figure 3a shows the transient magnetic field response, t-MPA(*B*) of the two PA bands at a fixed time *t*=200 ps; whereas Fig. 3b depicts the t-MPA time-evolution at a fixed field of 300 mT. The t-MPA experiment was also attempted on the $PA_{SE}$ band in the DOO-PPV polymer; however we found null response at any delay time (see Fig. 2a inset) showing that $PA_{SE}$ is due to SE and thus is not susceptible to magnetic field. We thus conclude that the t-MPA obtained in PDTP-DFBT (as well as in other DA-copolymers [32]) is a unique feature of the primary photoexcitation in the DA-copolymer. Although this species has a predominantly spin singlet character (since is instantaneously photogenerated) it is nevertheless a *composite particle* whose different components may have S≠0, hence the name CME. Importantly, the t-MPA(*B*) response seen in Fig. 3a cannot be understood using the 'Merrifield model' spin Hamiltonian of TT-pair (see S.I.) which describes well the intermolecular SF and triplet-triplet annihilation processes in various organic compounds[33], because this model does not fit the experimental t-MPA(*B*) response (see Fig. 4a). Consequently, the primary photoexcitation species in the copolymer is not a TT-pair per se. The t-MPA(*B*) response does not originate from a TE model with larger ZFS parameters either (see Fig. 1f), since such a species would not show *two* PA bands[25]. We thus conclude that the obtained t-MPA(*B*) response here needs be described by a novel spin-Hamiltonian that has not been used before in the field of 'organic magnetic field effect' (see below).

It is clear from Fig. 3a that the t-MPA(*B*) response of $PA_1$ and $PA_2$ have opposite polarity, but otherwise have the same shape (in contrast to the Merrifield TT-model[33], see S.I.). This shows that the two PA bands are *correlated*; namely, the field-induced population change in one CME spin component comes at the expense of population change in another spin component. Also the t-MPA evolution at fixed field (Fig. 3b) shows that the magnetic field induces opposite changes in the decay dynamics of the two PA bands. At *B*=0 the population of each CME component decays exponentially, $N(t) \sim \exp(-\nu_0 t)$, where $\nu_0$ is the recombination rate at *B*=0. A field-induced change, $\Delta\nu$ in the recombination rate changes the transient decay to $N_B(t) \sim \exp[-(\nu_0 \pm \Delta\nu)t]$. Since MPA=[PA(*B*)-PA(0)]/PA(0), we consequently obtain t-MPA(t)=A[1-exp(-$\Delta\nu$t)], where A is a constant that depends on spin-related parameters. This form perfectly describes the obtained t-MPA evolution in Fig. 3b using $\Delta\nu = 10^{10}$ s$^{-1}$. This indicates that the field increases the decay rate of PA$_1$, but decreases the rate of PA$_2$ by the same amount; with no field-induced change of their initial population. This explains the curious t-MPA evolution where t-MPA(0)=0, increases with time, then reaches saturation at long time.

To understand the genuine t-MPA(*B*) response in PDTP-DFBT we realized that there are four spin ½ particles that are involved in the CME species, since the $2^1A_g$ component in the copolymer contains in fact two 'electrons' and two 'holes' (see Fig. 4b inset). The four spins are split into two e-h pairs of which total spin should be calculated according to the Quantum Mechanic theory of addition of four spin ½ angular momenta; it thus contains one quintet, three triplets and two singlets, altogether 16 states (5+3*3+2*1=16) (see Methods and S.I.-S4). Therefore, the spin-Hamiltonian should act in a Hilbert space of $2^4$=16 dimensions that describes the spin interactions among the four spin ½ particles via several anisotropic exchange couplings and a Zeeman term, as detailed in the *Methods* and S.I. section. The full spin-Hamiltonian of the four spins in a magnetic field B||z is given by:

$$H = \sum_{i=1}^{4} g_i \mu_B B S_{iz} + H_{x1}(\theta_1,\varphi_1) + H_{x2}(\theta_2,\varphi_2) \quad (1)$$

where $g_i$ (≈2) are the gyro-magnetic parameter of each spin ½ particle with spin $S_i$, and the two other terms describe the anisotropic interaction between the spin ½ particles in the two e-h pairs, where the principal z-axis of each pair makes ($\theta,\varphi$) spherical angles with **B** (see S.I-S4). As a result, the 16 eignstates do not have the same energy, as seen in Fig. 4b inset. In the present case where the isotropic component of the spin exchange, J>~0.1 eV, is much larger than the anisotropic components, *D* and *E* (<~10 μeV), the 16 eignstates are arranged into several groups having different energies. These are: (i) the lowest energy group of nine states at $E_{CME}$~$2E_T$, which is equivalent to TT-pair (TT in Fig. 1d), where each sublevel contains a contribution from a spin quintet (Q), triplet (T$_1$) and singlet (S$_1$); (ii) two additional triplets (T$_2$ and T$_3$) at higher energy, E= $2E_T$+J; and (iii) a singlet state (S$_2$) at even higher energy, E=$2E_T$+2J.

The CME state has therefore multiple spins (namely 0, 1 and 2) that must be conserved in any optical transition. Consequently the CME characteristic PA bands are superposition of PAs in each manifold, namely the singlet (S), triplet (T), and quintet (Q) manifolds. The strongest transition in the singlet manifold is $^1$CME→'m$^1$A$_g$' at photon energy $E_1$≈0.4 eV[18] analogous to the transition $1^1B_u \rightarrow m^1A_g$ in more traditional PCPs. Therefore $PA_1$ has two contributions: a component that comes from the SE (namely '$1^1B_u$' $\rightarrow$ '$m^1A_g$') and another component from the CME ($^1CME \rightarrow$ '$m^1A_g$'), which is allowed due to the inversion symmetry breaking in DA-copolymers[18]. The strongest transition in the CME triplet manifold is from TT, namely TT$\rightarrow$TT$^*$, as depicted in Fig. 1d. The photon energy associated with this transition was recently calculated[18] to be slightly lower than $PA_T$ (see Fig. 1d). We thus identify $PA_2$ at 0.82 eV with this transition in the CME triplet manifold. Finally, transitions in the quintet manifold should be at much higher energies[18]. We therefore conclude that this simple 'four spin ½ model' is capable of explaining the two transient PA bands related with the CME species. In the following we show that this model also explains the t-MPA(*B*) response.

At time *t*=0 there is a strong optical transition from the ground state into the SE level, that is immediately followed by populating the CME singlet level, $S_1$ due to the strong SE-TT interaction. However the $S_1$ in the CME has contributions from all nine TT sublevels. The singlet and triplet characters of the nine lowest CME sublevels change with the field **B** (see Fig. 4b insets). If the CME singlet decay rate is different from that of the triplet, then the PA decay from the singlet component ($PA_1$) and triplet component ($PA_2$) would be B-dependent, and thus show t-MPA. To calculate the MPA(B, *t*) response we computed the spin eignfunctions and eignenergies of each of the 16 spin configurations as a function of *B*. In particular, we calculated the singlet and triplet contributions of the nine CME state (Fig. 4b insets) in order to obtain the recombination of each PA band (see S.I.-S4).

Figure 4b shows a theoretical calculation of the t-MPA(*B*) for $PA_1$ (singlet) and $PA_2$ (triplet) components of the CME at *t*=200 ps, using the exchange parameters $J_1 = J_2 > 0.1$ eV, $D_1 = D_2 = 45$ mT and E=0, and $S_1$ and $T_1$ recombination rates $\nu_1 = 2 \times 10^{10}$ s$^{-1}$ and $\nu_2 = 10^9$ s$^{-1}$, respectively. Not only that the t-MPA(*B*) response is reproduced by the theory, but we also correctly obtained opposite response for the t-MPA$_1$ and t-MPA$_2$ bands, in contrast to the simple TT-model (S.I. S5). This shows that the 16 dimension Hilbert space is necessary for dealing the 'four spin ½ particles' problem. In contrast, when we consider only nine TT spin states, as in the traditional Merrifield model[11,33], we cannot reproduce the experimental t-MPA response (see Fig. 4a and S.I.-S5).

It is interesting to study the CME photoexcitation decay at later time domain. Figures 3c and 3d show the t-PA decay and t-MPA evolution of the CME species in the microsecond (μsec) time domain (see *Methods* and S.I.-S2) and 40K. Figure 3c shows that the PA decay at 0.9 eV (where both $PA_2$ and $PA_T$ contribute) is strongly magnetic field dependent. From the change, $\Delta PA(t)$ in t-PA with $B$ we obtain the t-MPA($B,t$) response and study its time evolution. Figure 3c inset shows that the t-MPA at $B$=180 mT changes polarity at $t\sim4$ μsec. This is reflected in the t-MPA($B$) response (Fig. 3d), which dramatically changes for $1<t<10$ μsec. In fact t-MPA($B$) response changes from an early time line-shape that is similar to that obtained for the CME in the ps time domain (Fig. 3a), to a longer time line-shape similar to that of individual, uncorrelated triplets as in ss-MPA (Fig. 1f). We therefore interpret this interesting t-MPA($B$) evolution as *decomposition of the CME species* into two separated triplets having loosely or uncorrelated spins. This experimental result is a strong evidence for intrachain singlet fission (SF) process in the PDTP-DFBT chains, which indirectly also supports our CME interpretation. We emphasize that the SF process in the DA-copolymer is very much delayed compared to the ultrafast intermolecular SF process known to exist in some acene molecules such as pentacene and other molecules[11-13,34-36], since the CME decomposition process here is *intrachain*, and thus more difficult to achieve.

We now determine whether the CME photoexcitation can be ionized in donor-acceptor (D-A) blend, thereby generating electron-polaron and hole-polarons that contribute to the photocurrent in OPV devices. In order to identify the charge excitations in PDTP-DFBT, we first measured the charge polaron absorption spectrum by doping a pristine film with a strong acceptor, $HAuCl_4$ (see S.I.-S1) as shown in Fig. 5a. We identify two broad doping-induced absorption (DIA) bands, $P_1$ and $P_2$, due to polarons that peak, respectively, at 0.35 eV and 1.2 eV (see Fig. 5c); these DIA bands are typical to polarons in PCPs[37]. Figure 5a also displays the ss-PA spectrum of a PDTP-DFBT/$C_{71}$-PCBM (D-A) blend that shows maximum OPV efficiency[17]. The ss-PA spectrum contains the two polaron PA bands $P_1$ and $P_2$, similar to the DIA bands; and a third PA band due to triplets (i.e. $PA_T$ at 0.95 eV) that we identify from its MPA($B$) response (Fig. 1f). That $PA_T$ exists in the ss-PA of D-A blend shows that the TE is stable in the PDTP-DFBT copolymer chain, indicating that a direct TE dissociation into e-h polaron pairs (PP) across the D-A interfaces is *unlikely* in the copolymer/fullerene blend[38]. In this case the belated SF of the CME state into two TE needs be treated as a 'loss mechanism' for photovoltaic applications. To check whether there is another mechanism in which the CME directly ionizes into PP across the D-A interfaces in the copolymer/fullerene blend, we studied the ps t-PA dynamics in the DA-copolymer/fullerene blend.

The ps t-PA spectra in the D-A blend excited at 1.55 eV is shown in Fig. 6a. At $t=0$ the t-PA spectrum is similar to that in the pristine copolymer (Fig. 2b) which shows the two instantaneously generated $PA_1$ and $PA_2$. At $t>0$ the spectrum evolves, showing decay and red-shift of $PA_1$ and a blue-shift of $PA_2$. We also used a non-linear crystal and optics (see S.I.-S2) for obtaining a weak probe beam at 1.24 eV in order to monitor the evolution of the polaron band $P_2$ in the ps time domain (see Fig. 5a). As seen in Fig. 6b, simultaneously with the $PA_1$ and $PA_2$ decays there is a fast increase of about 2ps in the polaron PA band in the copolymer. This shows that the CME species may dissociate into polaron-pairs at the D-A interfaces in ~2ps. Consequently, the transient spectral shifts seen in the PA bands reflect the CME dissociation into polaron pairs. Under these conditions two of the CME components, namely the SE and TT-pair may dissociate into PP following two distinctive processes.

We interpret $PA_1$ red-shift as due to CME dissociation of the SE component into PP at the D-A interfaces, namely $PA_1 \rightarrow P_1$, similar to many other PCP/fullerene blends. The SE dissociation dynamics is shown in Fig. 6c at probe energy of 0.4 eV. We notice that the low-energy t-PA shows a fast decay into a plateau, because $P_1$ transition is close to that of $PA_1$, and the resulting polarons are long-lived in the blend. If the TT-pair component of the CME would also simply dissociate into PP, then the transient blue shift seen in the high energy PA band would be impossible to explain, since no polaron band overlaps with $PA_2$ (see Fig. 5a). We thus interpret the observed blue-shift in the high energy t-PA band as dissociation of the TT-pair component that follows the *unique reaction*:

$$TT_{S1} \rightarrow {}^{\uparrow}PP_T + {}^{\downarrow}TE, \qquad (2)$$

where $TT_{S1}$ is a TT-pair state in the singlet spin state ($S_1$; see Fig. 4b inset), ${}^{\uparrow}PP_T$ is a PP across the D-A interface having spin-up triplet configuration, whereas ${}^{\downarrow}TE$ is a spin-down TE in the copolymer (alternatively, ${}^{\downarrow}PP_T + {}^{\uparrow}T$ and ${}^{0}PP_T + {}^{0}T$ are also viable reactions). Reaction (2) is spin-allowed since the left and right hand sides both have total spin S=0. Energetically, this reaction is exothermic since $E_{TT} > E_T + E(PP_T)$ (Fig. 5c; see below). We thus interpret $PA_2$ blue shift as due to reaction (2), where TT→TE occurs with a unique spectral-shift in the high energy t-PA from 0.8 to 0.95 eV. We note that other reactions similar to reaction (2) may also occur if the TT is initially in the triplet ($T_1$) or quintet (Q) configurations. In fact *all nine* possibilities of $PP_T+TE$ spin addition can be matched to the nine spin configurations of the initial TT-pair state. This shows that the reaction TT→$PP_T$+TE is very probable, since there is no spin barrier or energy barrier to overcome. In fact, the interesting reaction TT→$PP_T$+$PP_T$ for all nine spin configurations is also viable if *hot TT* are initially photogenerated. This reaction may lead to *two e-h pairs* across the D-A interfaces. This may occur in DA-copolymers since the competitive SF reaction here is slow, in contrast to the fast SF reaction that occurs in the acenes[35].

We note that $PP_T$ is a PP where the electron resides on the $C_{60}$ molecule and the hole resides on the copolymer, mostly on the moiety with smaller electron affinity. The $PP_T$ energy, $E(PP_T)$ can be estimated from the emission spectrum of $PP_S$ at the D-A interfaces that may be excited in the copolymer/fullerene blend; this estimation rests on the assumption that $E(PP_T) < E(PP_S)$ by about 0.1 eV (Fig. 5c). Figure 5b shows the PL spectra of pristine and copolymer/fullerene blend films measured at ambient. The PL emission spectrum from the blend film is composed of a high energy PL band from the PDTP-DFBT copolymer with 0-0 emission at ~ 1.4 eV; and a broad PL band below 1.2 eV that is due to the $PP_S$ emission that peaks at $E(PP_S)$ ~0.95 eV (Fig. 5b inset). Since $E(PP_T)$ is within 0.1 eV of $E(PP_S)$[39], we estimate $E(PP_T)$ ~0.85eV (see Fig. 5c). Therefore, reaction (2) is energetically viable since $E_{CME} \approx E_T + E(PP_T) \approx 1.65$ eV.

The reaction TT→$PP_T$+TE is reflected in $PA_2(t)$ dynamics (Fig. 6b); it comprises of a fast decay into a plateau, similar to that of $PA_1(t)$ dynamics. However, in contrast to $PA_1(t)$, the plateau in $PA_2(t)$ is due to the longer-lived TE having $PA_T$ at 0.95 resulting from the TT dissociation (see reaction 2). In order to eliminate the plateaus from $PA_1(t)$ and $PA_2(t)$ dynamics at 0.82 eV and 0.43 eV, respectively (Fig. 6b), we calculate the time derivative of their respective decay kinetics, namely $d(PA_j(t))/dt$, j=1,2 (Fig. 6c). As is clearly seen that both PA(t) derivatives decay together, and exactly match $P_2(t)$ formation 'build-up'. This is strong evidence that both CME components, namely SE and TT-pair, may generate PP across the D-A interface.

We have followed $PA_T(t)$ dynamics in the PDTP-DFBT/fullerene blend to longer times, up to 1.2 ns (Fig. 6d) for studying the evolution of the TE species in the copolymer chains. Surprisingly, $PA_T$ shows a second rise at $t>20$ ps, reaching saturation at $t\sim800$ ps. At the same time $P_2(t)$ band decays with exactly the same dynamics to that of $PA_T(t)$ rise. This shows that the TE population in the copolymer chains increases *at the expense* of the PP species at the D-A interfaces, indicating that a 'back reaction' $PP_T \rightarrow TE$ occurs. This 'back reaction' is energetically possible in the PDTP-DFBT/$C_{71}$-PCBM blend since $E(PP_T)>E_T$; similar to other PCP/fullerene blends[40,41]. This reaction has been recognized as a 'photocurrent loss' mechanism in OPV solar cells, which has to be eliminated in order to increase the solar power efficiency beyond 10%.

We also measured the t-MPA($B$) response of $PA_T$ up to 500 ps, during which the 'back reaction' occurs (Fig. 6d). The negative t-MPA($B$) response measured at $t=500$ ps (Fig. 6d inset) is broad and unsaturated up to 300 mT; but is significantly narrower in this $B$-interval than the t-MPA($B$) response of the CME in the pristine film (Fig. 3a). We thus conclude that the t-MPA response of $PA_T$ in the blend at $t<\sim 800$ ps originates from a different spin-mixing process than that in the pristine copolymer. The spin-mixing in the blend occurs between $PP_T$ and $PP_S$ states at the copolymer/fullerene interfaces, that is probably mediated by the difference in the $g$-factor of electron and hole polarons (so called '$\Delta g$ mechanism'[42,43]) between the copolymer ($g\approx 2.002$) and the fullerene ($g\approx 1.998$). The conversion of the initially populated $PP_T$ (Eq.(2)) into $PP_S$ that increases upon the application of the magnetic field, reduces the population of $PP_T$ available for the 'back reaction'; and this, in turn decreases $PA_T$.

## *Methods*

**Picosecond pump-probe correlation spectroscopy;** The transient picosecond experimental setup is described in more detailed elsewhere[37]; it is a version of the well-known pump-probe correlation spectroscopy. The pump excitation beam was delivered by a fs Ti:sapphire laser; it was composed of pulses 150 fs duration, 0.1 nJ/pulse, 80 MHz repetition rate at 1.55 eV photon energy. A pump excitation at 3.1 eV was generated by doubling the 1.55 eV laser beam using a second harmonic generation crystal. The photoexcitations density (initially at $2\times 10^{16}$ cm$^{-3}$) was monitored by the changes, $\Delta T$ of the probe transmission, T (i.e. PA) induced by the modulated pump, measured by an InSb detector (Judson IR) using a phase-locked technique with a lock-in amplifier (SR830). The initial probe spectral range from 0.55 eV to 1.05 eV was supplied by an OPO Ti:sapphire based laser system (Spectra Physics). The probe spectral range was extended from 0.25 eV to 0.43 eV by a NLO crystal ($AgGaS_2$) using 'differential frequency' set-up. The pump beam was modulated at frequency of 50 kHz, and the PA (= $-\Delta T/T$) was measured using the lock-in amplifier set at the pump modulation frequency. A mechanically-delayed translation stage was introduced to the probe beam for measuring the PA at time, $t$ follow the pump pulse. The t-PA spectrum was constructed from about ~50 different wavelengths. For the transient polarization memory (POM) study we measured $\Delta T(t)$ where the pump/probe polarizations were parallel, $\Delta T_\parallel(t)$ or perpendicular, $\Delta T_\perp(t)$ to each other; the transient POM ($P(t)$) was calculated using the relation: $P(t) = (\Delta T_\parallel - \Delta T_\perp)/(\Delta T_\parallel + \Delta T_\perp)$.

**Transient μs to ms PA spectroscopy**: The optical set-up was the same as the cw PM apparatus except that the laser was pulsed. The pulsed pump excitation was an OPO laser (Quanta-ray) having 10 ns pulse duration at 10 Hz repetition rate. The OPA pump operated at 680 nm. The probe beam was an incandescent Tungsten/Halogen lamp set at 1 kW power. For monitoring $\Delta T(t)$ we used a laser diode at 1300 nm. The t-PA was monitored using a fast InGaAs detector (Thorlabs), coupled to a data acquisition card ATS9462 with 100 MHz bandwidth. A potentiometer was set to 1 kΩ to establish the detector gain. The time response of this set up was <0.5 μsec.

**Time dependent MPA spectroscopy:** We used the same setups described above except for the electro-magnet. The samples were mounted in a variable temperature cryostat and placed in between the two poles of bipolar electro-magnet. The t-MPA($B$) response is defined as t-MPA(%) = [PA(t,$B$)- PA(t,0)]/PA(t,0), where PA(t,$B$) is the PA(t) at field $B$. t-MPA($B$) in the μsec time domain was measured using the same electromagnet as in the ss-MPA (S.I. Methods). This response was compiled from PA($B$,t) dynamics at about 100 different field values from -180 to 180 mT. In the ps time domain there is a complication due to the background PA (see S.I. Methods). Under these conditions the t-MPA($B$) was obtained by subtracting the MPA($B$) response of the background PA that was measured separately at $t$=-10 ps.

**Four spin ½ Hamiltonian and MPA calculation**:

We use the straightforward basis of the four S=½ system:

$$\phi = |S_{1z}S_{2z}S_{3z}S_{4z}> , \quad S_{iz} = \pm\frac{1}{2} \tag{3}$$

According to the rule of the addition of angular momenta, the sum of four S=1/2 spins have maximum angular momentum J=2 and minimum J=0; In order to write the |J,M > wave functions we note that the final species are obtained as follows: (we denote by $^mL_i$ angular momentum with multiplicity m of particle i):

$$\begin{aligned} ^2L_1 + {}^2L_2 &\to {}^1L_{12} + {}^3L_{12} \\ ^2L_3 + {}^2L_4 &\to {}^1L_{34} + {}^3L_{34} \end{aligned} \tag{4}$$

In order to obtain the final configuration we add the four angular momenta obtained in (2):

$$\begin{aligned} ^3L_{12} + {}^3L_{34} &\to {}^5L + {}^3L + {}^1L \equiv Q + T_1 + S_1 \\ ^1L_{12} + {}^1L_{34} &\to {}^1L \equiv S_2 \\ ^1L_{12} + {}^3L_{34} &\to {}^3L \equiv T_2 \\ ^3L_{12} + {}^1L_{34} &\to {}^3L \equiv T_3 \end{aligned} \tag{5}$$

All 16 wave functions for the four spin wavefunctions in (5) can be written as various combinations of (3). Furthermore, we introduced anisotropic exchange interaction (AXI) between spins 1 and 2 ("first pair") and another AXI between spins 3 and 4 ("second pair"):

$$\begin{aligned} H_{x1} &= E_{T1} - J_1(\vec{S}_1 \cdot \vec{S}_2 - 1/4) + \vec{S}_1 \cdot \tilde{V}_1 \cdot \vec{S}_2 \\ H_{x2} &= E_{T2} - J_2(\vec{S}_3 \cdot \vec{S}_4 - 1/4) + \vec{S}_3 \cdot \tilde{V}_2 \cdot \vec{S}_4 \end{aligned} \tag{6}$$

where $J_i$ are the isotropic exchange and $V_i$ are 3x3 traceless tensors describing the anisotropy. The additive term J/4 in Eq. (6), assures that for isotropic exchange the triplet state of each of $H_{x1}$ and $H_{x2}$ is at energy $E_T$ while the singlet state is at $E_S = E_T + J$. In the principal frame of reference of each pair $V_i$ is diagonal with two independent anisotropy parameters, $D_i$ and $E_i$ (denoted below as ZFS parameters),

$$H_{x1} = E_{T1} - J_1(\vec{S}_1 \cdot \vec{S}_2 - 1/4) + \frac{D_1}{3}[2S_{1z}S_{2z} - \frac{1}{2}(S_{1+}S_{2-} + S_{1-}S_{2+})]$$

$$+ \frac{E_1}{2}(S_{1+}S_{2+} + S_{1-}S_{2-})$$

$$H_{x2} = E_{T2} - J_2(\vec{S}_3 \cdot \vec{S}_4 - 1/4) + \frac{D_2}{3}[2S_{3z}S_{4z} - \frac{1}{2}(S_{3+}S_{4-} + S_{3-}S_{4+})] \quad (7)$$

$$+ \frac{E_2}{2}(S_{3+}S_{4+} + S_{3-}S_{4-})$$

Since the interactions are anisotropic, the relative orientation of the two pairs affects the energy levels, wave functions, and spin configurations. Furthermore, the orientation of the external magnetic field, **B**, with respect to the pairs is also critical. In the laboratory frame of reference with B∥z, the z-axis of each pair makes (θ,φ) spherical angles with B; the tensor $\tilde{V}$ (for each pair) has 5 components as given in the Suppl. Info.

For the t-MPA of the T and S components of the CME species the two spin pairs experience negative exchange and very large J (>0 in Eqs. (6), and(7)), and finite, but small D and E: |J|>>D>E. In this case we obtain three groups of levels: at ~2$E_T$ (CME=$S_1$+$T_1$+Q, 9 levels), at ~2$E_T$+J=$E_T$+$E_S$ ($T_2$,$T_3$, 6 levels), and at 2$E_T$+2J=2$E_S$ ($S_2$, single level).. The lowest group is the CME complex (first line in Eq. (5)). An enlarged view of the lowest group is shown in the S.I.

 Acknowledgments

The work at the University of Utah was supported by the DOE grant No. DE-FG02-04ER46109 (ps transient spectroscopy; UNVH); and NSF-MRSEC program DMR 1121252 (cw spectroscopy; TPB as well as the acquisition of the experimental facilities for the μsec transient spectroscopy). Work at Arizona was partially supported by NSF grant DMR-1151475, and the UA-REN Faculty Exploratory Research Grant. The MPA calculations were supported by the Israel-USA BSF (Grant # 2010135; EE and ZVV). The work at UCLA was supported by ONR (N00014-14-1-0648), and AFOSR (FA9550-12-1–0074).

**Author contributions** U.N.V.H. measured the transient PA and MPA spectroscopy. T.P.B. measured the steady state spectroscopies. L.D., G.L. and Y.Y. designed and synthesized the PDTP-DFBT copolymer. S.M. and Z.V.V. conceived the idea of SE-TT interaction in the PCC; K.A. calculated the PA bands from the SE and TT-pair; E.E. designed the four spin ½ model, and calculated the ss-MPA(*B*) and t-MPA(*B*) responses; and Z.V.V. designed the project and wrote the first draft.

**Competing financial interests** The authors declare no competing financial interests.

**Materials & Correspondence** Correspondence and requests for materials should be addressed to Z.V.V. (val@physics.utah.edu).

# Figures and Legends

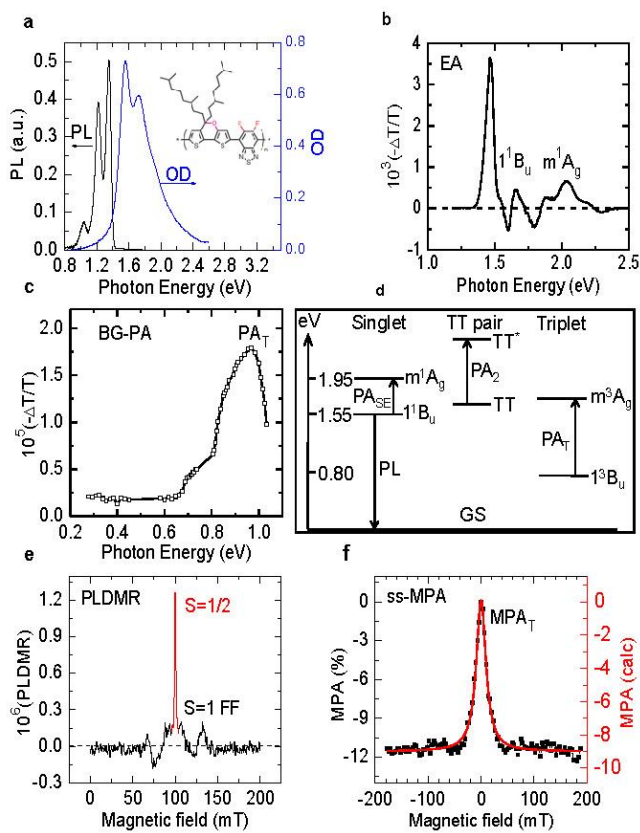

**Figure 1 | Steady state (ss) spectroscopies of pristine PDTP-DFBT π-conjugated copolymer** (repeat unit is shown in panel **a** inset). **a,** The photoluminescence (PL) and absorption spectra of the copolymer film. **b,** The electroabsorption (EA) spectrum, where the two important excited states in the singlet manifold are assigned. **c**, The steady state photoinduced absorption (ss-PA) spectrum measured via the background PA (BG-PA) in the picosecond pump-probe correlation, modulated at 1 kHz (see S.I.). The triplet PA (PA$_T$) is assigned. **d**, Schematics of the main energy levels and associated optical transitions in three different manifolds of the copolymer, namely: singlet exciton (SE), TT-pair, and isolated triplet (TE), respectively; without considering possible interaction between SE and TT-pair. **e**, The PL-detected magnetic resonance, PLDMR(*B*) response of the copolymer measured at 10K. The full-field (FF) triplet powder pattern (black) and spin ½ resonance line (red) are assigned. **f**, The steady state magneto-PA (ss-MPA(*B*)) response of PA$_T$ band measured at 40K. The line through the data points is a fit based on individual triplet exciton using the zero-field splitting parameters *D*=38 mT and *E*=15 mT (see model in S.I.).

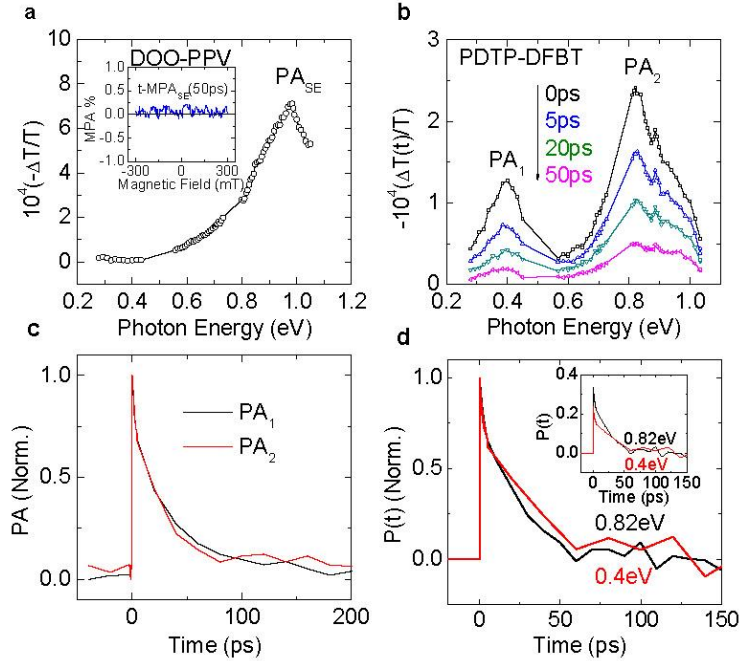

**Figure 2 | Room-temperature ps transient PA spectroscopy of pristine PDTP-DFBT embedded in a polystyrene matrix**. **a,** The transient PA (t-PA) of DOO-PPV polymer film in the mid-infrared measured at $t$=0 excited at 3.1 eV. Only a single PA band of the singlet exciton ($PA_{SE}$) is observed. The inset shows the lack of transient magneto-PA (t-MPA($B$)) response at 100 ps. **b,** The time evolution of the t-PA spectrum in PDTP-DFBT measured at several delay times, $t$ following the pump excitation at 1.55 eV. The t-PA bands of the CME species, $PA_1$ and $PA_2$ are assigned. **c,** The decay dynamics of $PA_1$ (red line) and $PA_2$ (black line) up to 200 ps. **d,** The decay dynamics of the normalized polarization memory, P(t) for $PA_1$ and $PA_2$. The inset shows P(t) of the two bands without normalization; the P(t=0) value of $PA_1$ and $PA_2$ are 0.2 and 0.35, respectively.

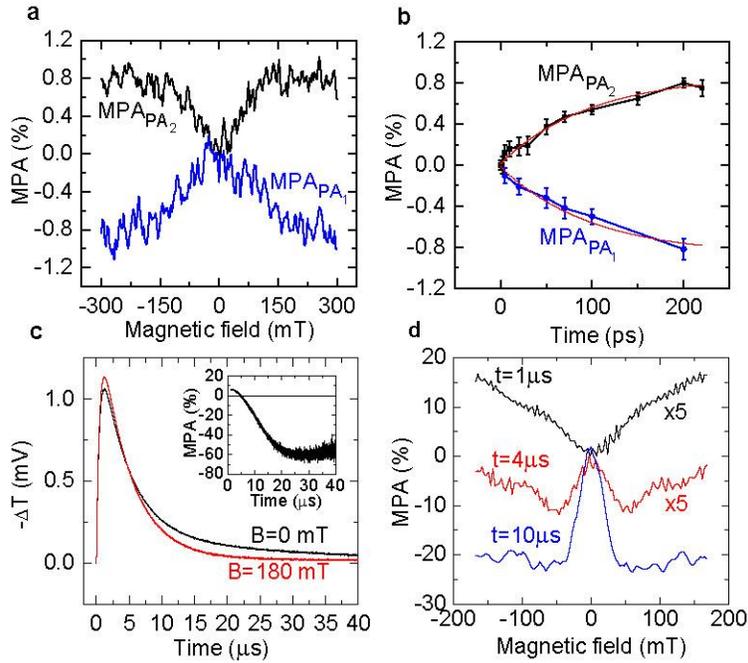

**Figure 3 | Transient magneto-PA (t-MPA) response of pristine PDTP-DFBT film in the ps to microsecond time domains. a**, The t-MPA($B$)) response of PA$_1$ (blue line) and PA$_2$ (black line) measured at $t$=200 ps up to $B$=300 mT. **b**, The evolution of the t-MPA($B$=300 mT) for PA$_1$ (blue line) and PA$_2$ (black line) up to $t$=200 ps. The red lines through the data points is a fit based on the magnetic field change in the recombination rates (see text). **c**, The PA decays in the μsec time domain measured at 0.9 eV and 40K at magnetic field $B$=0 (black line) and $B$=180 mT (red line), respectively up to $t$=40 μsec. The inset shows the t-MPA at $B$=180 mT up to 40 μsec calculated from the PA decays dependence on $B$. **d**, The t-MPA($B$) response up to $B$=180 mT measured at different times, $t$ as indicated.

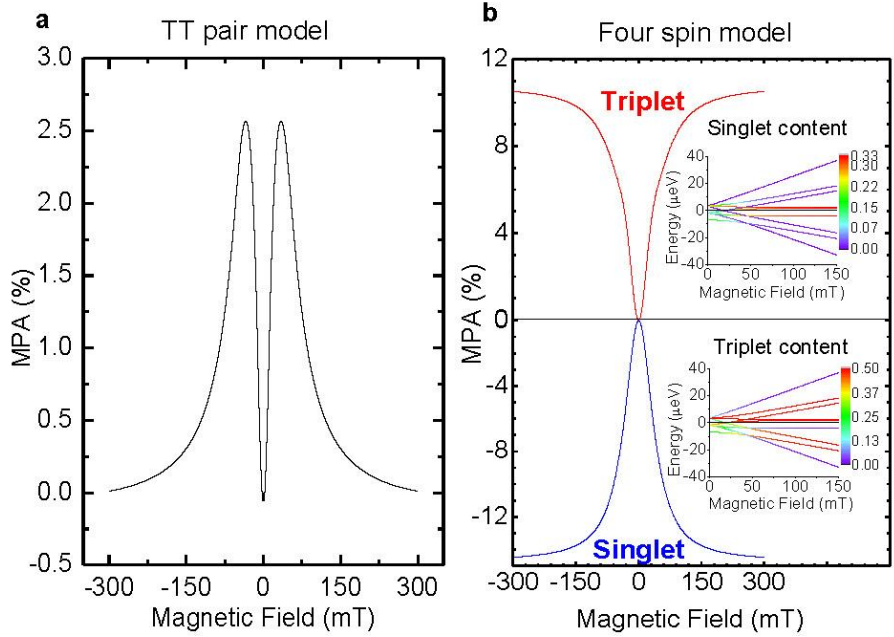

**Figure 4 | Theoretical model calculations of the t-MPA(*B*) response. a**, Theoretical calculation of the t-MPA(*B*) response at *t*=200 ps based on the TT-pair 'Merrifield model'[33] (see S.I.), where the spin of the two triplets are parallel to each other, using the same parameters as in (b). **b**, Theoretical calculation of the t-MPA(*B*) response at *t*=200 ps based on the 'four spin ½ model' using *D*=45 mT, E=0 and the recombination rates $\nu_1=2\times10^{10}$ s$^{-1}$ and $\nu_2=10^9$ s$^{-1}$, respectively, for the $S_1$ and $T_1$ spin states in the CME lowest level. We note that the spin triplet and singlet response have opposite polarity, and do not show the intricate shape as in (a); in agreement with the data in Fig. 3a. The insets show in 'false colors' the magnetic field dependence of the energy levels and singlet and triplet component weights of the nine TT states using the 'four-spin ½ model'; similar plots for the 'Merrifield model' are given in the S.I. Fig. S1.

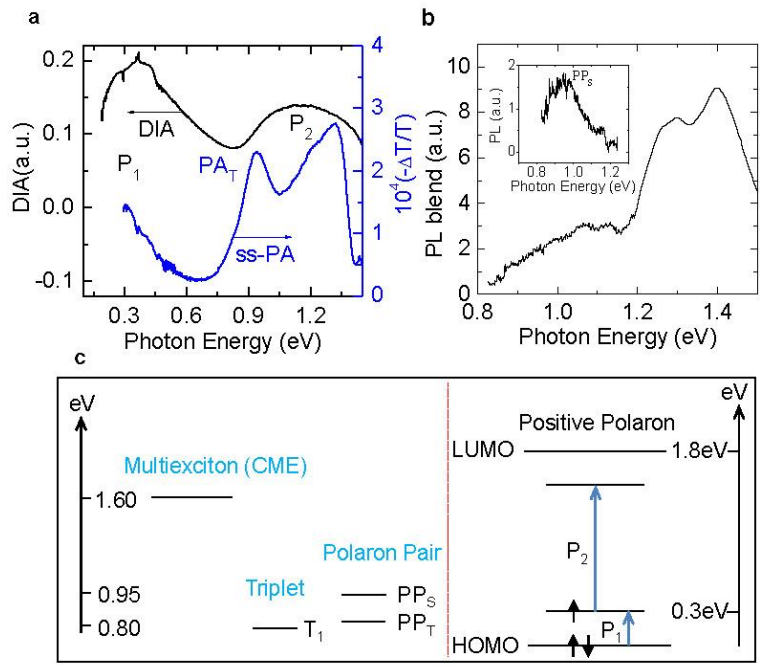

**Figure 5 | Steady state spectroscopies of PDTP-DFBT/$C_{71}$-PCBM blend film. a**, Doping induced absorption (DIA) spectrum (black line) of pristine PDTP-DFBT, and steady state PA (ss-PA) spectrum (blue line) of the PDTP-DFBT/$C_{71}$-PCBM blend measured at 80K. The DIA and PA bands $P_1$ and $P_2$ of polarons in the copolymer chains, and the PA band of triplet ($PA_T$) are assigned. **b**, The photoluminescence (PL) emission spectrum of the PDTP-DFBT/$C_{71}$-PCBM blend film excited at 680 nm. The PL bands of the pristine copolymer and singlet polaron pair, $PP_S$ at the D-A interfaces are assigned. The inset shows the PL from $PP_S$ after the PL spectrum from the pristine copolymer is subtracted out, showing a $PP_S$ peak at 0.95 eV. **c**, Left panel: energy levels of various photoexcitation species in the DA-copolymer chains and copolymer/fullerene interfaces; right panel: energy levels and optical transitions of positive polarons in the DA-copolymer chains.

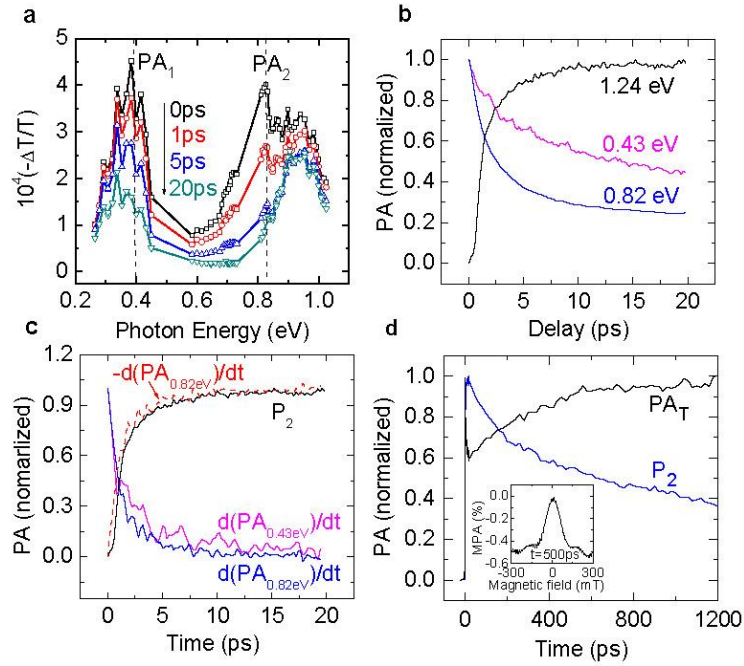

**Figure 6 | ps transient PA spectroscopies of PDTP-DFBT/$C_{71}$-PCBM blend film. a**, The evolution of the t-PA spectrum in the blend at various times, $t<20$ ps. The PA bands $PA_1$ and $PA_2$ are assigned for $t=0$; at $t>0$ these PA bands gradually transform into $P_1$ (polarons) and $PA_T$ (triplets), respectively. **b,** The t-PA dynamics up to 20 ps measured at three different photon energies as indicated, that represent the bands $P_2$ (black line), $PA_1$ (red line), and $PA_2$ (blue line), respectively. **c**, Comparison between the t-PA rise dynamics at 1.25 eV ($P_2$; black line) and the time derivatives of the t-PA decay dynamics at $PA_1$ (purple line) and $PA_2$ (blue line). Also the polarity reversed $d(PA_{TT})/dt$ dynamics (red broken line) is compared to $P_2(t)$ dynamics (black line). **d**, Transient sub-nanosecond dynamics of the triplet PA band ($PA_T$) and polaron PA ($P_2$) in the copolymer/fullerene blend. The inset shows the t-MPA($B$) response of the $PA_T$ band at $t=500$ ps.

Supplementary Information

# Novel primary photoexcitations in π-conjugated donor-acceptor copolymers probed by transient magneto-photoinduced-absorption

Uyen N. V. Huynh[1], Tek P. Basel[1], L. Dou[2], Karan Aryanpour[3], Gang Li[2], Sumit Mazumdar[3], Eitan Ehrenfreund[4], Yang Yang[2], and Z. Valy Vardeny[1*]

[1]*Department of Physics & Astronomy, University of Utah, Salt Lake City, Utah 84112*

[2]*Department of Materials Science & Engineering, University of California-Los Angeles, Los Angeles, California 90095*

[3]*Department of Physics, University of Arizona, Tucson Arizona 85721*

[4]*Department of Physics, Technion-Israel Institute of Technology, Haifa 32000, Israel*

*\*Correspondence to:  val@physics.utah.edu*

# I. Supplementary Methods

## S1: The PDTP-DFBT DA-copolymer pristine, doped and blend films

(i) Pristine films: The DA-copolymer was synthesized at the University of California Los Angeles. The synthetic route, chemical structure, BHJ solar cell device fabrication and PCE measurements were described in ref. 17[17]. Neat films were prepared by drop cast or spin coating from a solution of pristine PDTP-DFBT (or PDTP-DFBT/PC$_{71}$BM blend with mixing ratio 1:2 by weigh) dissolved in dichlorobenzene (7mg/ml) on CaF$_2$ substrates for t-PA measurements, and on sapphire substrates for all other optical measurements.

(ii) Doped films: For the doping induced absorption measurements, a pristine PDTP-DFBT film was doped with HAuCl$_4$, which is known to be a strong acceptor. The HAuCl$_4$ powder was first dissolved in acetonitrile at 0.01M concentration, and stirred overnight to mix uniformly. The film was then dipped in the solution for ~1 minute.

(iii) D-A Blend films: The PC$_{71}$BM, [6,6]-Phenyl C$_{71}$ butyric acid methyl ester > 99% fullerene powder were bought from Sigma Aldrich and used as received. All solutions and films were prepared in a glove box filled with N$_2$.

## S2: Optical spectroscopies

Various optical spectroscopies that include transient photoinduced absorption (t-PA), steady state photoinduced absorption (ss-PA), electro absorption (EA), and doping induced absorption (DIA) were used to study the short- and long-lived photoexcitations and the absorption of thin film PDTP-DFBT copolymer and its blend with PC$_{71}$BM.

**(i) Picosecond pump-probe correlation spectroscopy**: The transient picosecond experimental set-up was described in detailed elsewhere[37]; it is a version of the well-known pump-probe correlation spectroscopy. The pump excitation beam was composed of pulses 150 fs duration, 0.1 nJ/pulse, at 80 MHz repetition rate from a fs Ti:sapphire laser that was set at 1.55 eV photon energy. A pump excitation at 3.1 eV was generated by doubling the 1.55 eV beam using a second harmonic generation crystal. The initial photoexcited exciton density in the polymer film was adjusted to be ~$5 \times 10^{16}$ cm$^{-3}$/pulse. The photoexcited species were monitored by the changes, $\Delta T$ of the probe transmission, T (i.e. PA) that was produced by the pump excitation. The probe spectral range was extended from 0.55 eV to 1.05 eV that was generated from an OPO Ti:sapphire based laser from Spectra Physics that gives both 'signal' and 'idler' beams. We also extended the probe spectral range from 0.25 eV to 0.43 eV by phase matching the "signal" and "idler" beams in a differential frequency crystal (AgGaS$_2$).

The pump beam was modulated at frequency, $f$=50 kHz, and the PA (= $-\Delta T/T$) was measured using an InSb detector (Judson IR) and a lock-in amplifier (SR830) set at $f$. A translation stage was introduced to the probe beam that could delay the probe pulses mechanically (1 ps=300 µm mechanical delay) thereby measure the PA at time, $t$ set by the delay line. For each probe wavelength we swept the delay line back and forth several times until a reasonable S/N ratio was achieved. The t-PA spectrum was then constructed from the t-PA at ~50 different wavelengths. For a weak probe beam at 1.24 eV used to monitor the PP dynamics in the PDTP-DFBT/C$_{71}$-PCBM blend, we used a double frequency crystal to generate the second harmonic from the 0.62 eV idler beam. We also used a 1300 nm 'short-pass filter' before the sample to block the 0.62 eV fundamental beam and a 'band-pass filter' centered at 1000±10 nm in front of InSb detector.

Since the pump and probe beams are linearly polarized we could also measure the polarization memory and its dynamics as a function of the probe photon energy. For the transient polarization memory we measured $\Delta T(t)$ where the pump/probe polarization were parallel, $\Delta T_{para}$ or perpendicular $\Delta T_{per}$ to each other. The polarization memory, P(t) is defined as: P(t) = $[\Delta T_{para}(t)-\Delta T_{per}(t)]/[\Delta T_{para}(t)+\Delta T_{per}(t)]$.

**(ii) <u>Transient PA spectroscopy in the µsec time domain</u>**: The pulsed excitation for this time domain was an OPO laser (Quanta-ray Indi model) operating at 10 Hz repetition rate having 10 ns pulse duration. The OPA pump at 355 nm 'center-wavelength' excited a basiScan OPO for generating pulses that are tunable across a broad spectral range from 410 nm to 2500 nm. The probe beam was an incandescent Tungsten/Halogen lamp at 1 kW power. For monitoring the transient ΔT we used several band pass filters on the probe beam as needed, or a laser diode with specific wavelength. The t-PA was monitored using a fast photodiode, namely InGaAs detector from Thorlabs, coupled to a data acquisition card ATS9462 with 100 MHz bandwidth. For this project the pump was set at 680 nm and ΔT(t) was measured at 1300 nm using a laser diode. This wavelength was chosen because it is possible to detect both triplet and triplet pair species in the PDBT-DFBT copolymer. A potentiometer was set to 1 kΩ to establish the detector gain. The time response of this set up was <0.5 µsec. The thin films were mounted in a closed cycle He refrigerator cryostat for low temperature measurements.

**(iii) <u>Background PA in the ps pump-probe measurements:</u>** When using pulses at 80 MHz repetition rate in the ps pump-probe experiment from a Ti-Sapphire laser as pump excitation, the time elapsed between successive pump pulses is ~12.5 ns. In this case some of the long-lived photoexcitations generated from one pulse do not completely recombine until the arrival of the next pulse, and thus contribute to a background PA signal[37]. In fact the transient PA rides on top of a 'background PA' as seen in Figure S1. The accumulation of the background photoexcitations from many pulses generates a steady state 'background PA'. This 'background

PA' is in fact modulated at frequency, $f$= 50 kHz, which is the pump modulation frequency in the ps set-up; and thus can serve as a convenient way for measuring the ss-PA at fast modulation frequency. We measured the 'background PA' in the present study at $f$=1 kHz. The pump in this case was the Ti-Sapphire laser beam, and the probe originates from the OPO 'signal' and 'idler' beams. This probe beam is much stronger than the probe beam from an incandescent light source, especially in the mid-IR spectral range. The combination of strong probe beam and fast modulation frequency (away from 1/$f$ noise) is ideal for measuring weak ss-PA that originates from photoexcitations that decay in the μsec time domain. As a matter of fact the 'background PA' in the pump-probe experiment is the only way of measuring weak ss-PA in the mid-IR, and we used it here for measuring the triplet PA spectrum at room temperature.

**(iv) Steady state spectroscopies**: For the ss-PA and PL we used a standard photomodulation set-up[27,44]. Thin PDBT-DFBT films were placed in a closed cycle He refrigerator cryostat operating at low temperatures. A 660 nm diode laser was used as a pump excitation and an incandescent tungsten/halogen lamp was used as a probe source. The pump beam was modulated at frequency $f$=310 Hz with a mechanical chopper. The changes of the probe transmission, ΔT induced by the laser pump excitation were measured using a monochromator, and various combinations of gratings, filters, and photodetectors spanning the spectral range 0.3< ℏω(*probe*) <2.3 eV. To increase the S/N ratio, the detector preamplifier was connected to a lock-in amplifier (SR830) referenced at $f$.

The absorption and DIA spectra were measured with UV-VIS-NIR absorption CARY 17 spectrophotometer and FTIR spectrometer, respectively. For the DIA measurement, pristine PDTP-DFBT film was dipped in the dopants solution for ~1min and the consequent doping induced absorption was measured.

The detail of the EA spectroscopy is described in reference[44]. Thin pristine PDTP-DFBT copolymer was deposited on an EA substrate in the form of an interdigitated gold electrode array pattern with a 40μm gap between the fingers electrodes. An AC electric field of the order of $10^5$ V/cm, at $f$=500 Hz modulation frequency was applied using a transformer and SR830 lock-in internal reference. The EA spectrum was measured at $2f$ since the polymers are randomly oriented in the film, and thus only a quadratic EA signal is generated.

## S3: Magneto photoinduced absorption spectroscopies

(i) <u>Steady state MPA</u>: The ss-MPA is defined by the relation, MPA(%) = (PA($B$)-PA(0))/PA(0), where PA($B$) is the PA at field $B$[29]. It shows the percentage change of the ss-PA under the influence of a magnetic field. For measuring the ss-MPA($B$) response, we used the same set-up as for the ss-PA experiment described above (section III (iv)) except for the magnetic field. The samples were mounted in the He cryostat and placed in between the two poles of bipolar electro-magnet. With the limit of 2.8 Amp feeding current and the gap between the poles of 5 cm, a maximum $B$ field that is achieved is ~180 mT (as measured by a magnetometer). For measuring the ss-MPA($B$) response we swept the feeding current of the magnet from 2.8A to -2.8A several times until a satisfactory S/N ratio was obtained.

(ii) <u>t-MPA in the μsec time domain:</u> The t-MPA($B$) response in the μsec time domain was measured using the same electromagnet as in the ss-MPA. This response was compiled from PA($B$,t) dynamics at about 100 different field values from -180 to 180 mT and therefore the S/N ratio is inferior to that of the ss-MPA.

(iii) <u>t-MPA in the ps time domain:</u> In the picoseconds time domain there is a complication due to the background PA. Under these conditions the t-MPA($B$) was obtained by subtracting the MPA(*B*) response of the background PA that was measured separately at *t*=-10 ps. This response is similar to the ss-MPA(*B*) response. The procedure to obtain the t-MPA is therefore the following: t-MPA(t,B)=[ΔPA(t,*B*)-ΔPA(t=-10ps,*B*)]/[(PA(t,*B*=0)-PA(t=-10ps,*B*=0)], where the terms ΔPA(t,*B*)= PA(t,*B*)-PA(t,*B*=0); PA(t=-10ps) is the background PA component; and PA(t) is the total PA signal, namely the summation of the transient PA *and* background PA.

## S4: <u>Calculation methods for the t-MPA(*B*) and ss-MPA(*B*) responses</u>

For the calculations of the MPA(*B*) response of either isolated TE or the CME we have calculated the time dependent population of the relevant levels using a simple TE spin Hamiltonian ($H_T$, in N=3 dimension Hilbert space) or a "four spin ½" Hamiltonian ($H_4$, N=16), respectively. Once the energy levels and wave functions of the relevant spin Hamiltonian, H, are calculated, the time dependent probability of finding the system in a given spin configuration can be determined using the density matrix approach. Assuming spin dependent recombination we then show that the level population becomes magnetic field dependent leading to MPA[29].

There are N different spin states ("spin configurations") in the spin Hamiltonian. We label these configurations by λ, where λ=$T_1$, $T_0$, $T_{-1}$ for the three triplet states of $H_T$. For $H_4$, λ takes 16 values representing the various singlets (two x 1 states), triplet (three x 3 states) and quintet (one 5 states) configurations. Assigning a spin dependent decay rate, $\kappa_\lambda$, to each spin configuration, the decay rate for each of the n=1,…,N levels becomes

$$\gamma_n = \sum_\lambda \kappa_\lambda P_{nn}^\lambda \quad (S1)$$

where $P_{nn}^\lambda$ is the $n^{th}$ diagonal matrix elements of the projection operator for the $\lambda$ spin configuration; note that $P_{nn}^\lambda$ is $B$-dependent. The time dependent probability for the system to be in the $\lambda^{th}$ spin configuration may now be written as[45]

$$\rho_\lambda(t) = Tr(P^\lambda \sigma(t)) = \sum_{n,m} P_{n,m}^\lambda \sigma_{m,n}(0) \cos(\omega_{mn} t) \exp(-\gamma_{mn} t) \ , \qquad (S2)$$

where $E_n = \hbar \omega_n$ (n=1,...N) are the energies of $H$, $\omega_{nm} = \omega_n - \omega_m$ ; $\gamma_{nm} = \gamma_n + \gamma_m$ and $\sigma(t)$ is the time dependent density matrix. $\sigma(0)$ is determined by the initial conditions; for optical excitation the system is initially in a singlet state: $\sigma(0)=P^S$. To calculate PA(t) we assume that the optical cross section is spin independent; consequently the time dependent absorption is obtained by integrating Eq. (S2)

$$PA_\lambda(t) \propto \int_0^t \rho_\lambda(t') dt'. \qquad (S3)$$

$PA_\lambda(t)$ thus obtained is magnetic field dependent since the probabilities $\rho_\lambda(t)$ are $B$-dependent. From Eq. (S3) we obtain the magneto-PA of a given spin configuration:

$$MPA_\lambda(B,t) = \frac{PA_\lambda(B,t) - PA_\lambda(0,t)}{PA_\lambda(0,t)} \ . \qquad (S4)$$

The time dependent response, t-MPA($B$) is given by Eq. (S4); the steady state MPA response, ss-MPA($B$) is obtained by letting $t \to \infty$.

### (i) MPA($B$) response related to isolated triplet excitons:

In pristine PDTP-DFBT films the steady state photoexcited TE density is low, and thus effects of TE-TE annihilation are small. In this case the TE density is determined by a non-radiative recombination process for which the spin sub-level recombination constants $\kappa_\lambda$ ($\lambda=\pm1,0$) are different from each other. The spin Hamiltonian for an isolated TE is determined by the two zero field splitting (ZFS) parameters, $D$ and $E$. In a magnetic field B||z making spherical angles $(\theta,\varphi)$ with the triplet principal axis, $H_T$ is given by[46]:

$$H_T = g\mu_B B S_z + \vec{S}\cdot\tilde{V}\cdot\vec{S} = g\mu_B B S_z + V_0[3S_z^2 - S(S+1)] + \\ V_{-1}(S_+ S_z + S_z S_+) + V_1(S_- S_z + S_z S_-) + V_{-2} S_+^2 + V_2 S_-^2 \quad (S5)$$

where S=1 is the TE spin and $\tilde{V}$ is a symmetric traceless tensor of rank 2 whose five independent elements are given by

$$V_0 = \frac{D}{6}(3\cos^2\theta - 1) - \frac{E}{2}\cos 2\varphi \sin^2\theta$$
$$V_{\pm 1} = \pm\frac{iD}{2}\sin\theta\cos\theta \pm \frac{iE}{2}\sin\theta\cos\theta\cos 2\varphi - \frac{E}{2}\sin\theta\sin 2\varphi \quad (S6)$$
$$V_{\pm 2} = -\frac{D}{4}\sin^2\theta - \frac{E}{4}(1+\cos^2\theta)\cos 2\varphi \mp \frac{iE}{2}\cos\theta\sin 2\varphi]$$

The principal ZFS parameters were obtained for TE in PDTP-DFBT from the PL detected magnetic resonance technique; D≈38 mT and E≈15 mT. Using these ZFS parameters we calculated the TE energy levels and wave functions in **B** applied in a general direction. We then calculated the ss-MPA(B) powder pattern using Eqs. (S2-S4) and averaging over all angles; as seen in the theoretical fit in Fig. 1f.

### (ii) **TT-pair model for MPA(*B*) response; Merrifield model**[33]

Two separated TE may combine to form a TT-pair when the TE density is not too small. TT annihilation is responsible for delayed fluorescence and shows itself in magneto-spectroscopy such as MPL, MEL and MPA of organic semiconductors[29].

The TT-pair Hamiltonian is constructed by adding $H_T$ (Eq. (S5)) for each of the two TE and a TE-TE interaction term. It is generally assumed that the interaction term is small compared to the ZFS of the TE and in this case the TT Hamiltonian is written as[33]

$$H_{TT} = H_{T1} + H_{T2} \quad (S7)$$

where $H_{T1}$, $H_{T2}$ are given by Eq. (S5) and the TE-TE interaction term has been omitted. The energies and wave functions of all the 9 states can be calculated for arbitrary spherical angles of each of the triplets. The calculated singlet and triplet content of each of the levels are shown in Fig. S1 for a magnetic field along the z-axis of the parallel pairs; these are in agreement with the classic Merrifield calculations[33]. Using Eqs. (S5)-(S7) we have calculated the level population densities and the powder pattern MPA response shown in Figs. 4a and S1. This TT model based on nine spin states does not fit the experimental t-MPA($B$) response that we obtained in DA-copolymers (Fig. 3 in the text).

### (iii) 'four spin ½ model' for t-MPA($B$) response

We realize that the TT-pair contains in fact four spin ½ particles, namely two electrons and two holes that are involved in the CME species. The Hilbert space in which the four spin ½ system is described has $2^4=16$ dimensions. Adding the angular momenta in pairs we may decompose it to a triplet-triplet pair ("TT pair") which further decomposes to a quintet (Q), triplet ($T_1$) and a singlet ($S_1$), two additional triplets ($T_2$ and $T_3$) and one additional singlet ($S_2$); altogether 16 states[47]. The wave functions for the 16 states may be written using the following basis for the four S=½ system:

$$\phi = |S_{1z}S_{2z}S_{3z}S_{4z}> \, , \, S_{iz} = \pm \frac{1}{2} \, . \tag{S7}$$

All 16 wave functions for the various configurations can be written as various combinations of (S7). For example, the singlet $|S_1>$ and triplet $|T_{1,0\pm}>$ are written below

$$\begin{aligned} |S_1> = &\frac{1}{\sqrt{3}}[|++--> + |--++> \\ &-\frac{1}{2}(|+-+-> + |+--+> + |-++-> + |-+-+>)] \end{aligned} \tag{S8}$$

$$|T_{10}> = \frac{1}{\sqrt{2}}[|++--> - |--++>]$$

$$|T_{1\pm}> = \frac{1}{2}[|\pm\pm+-> + |\pm\pm-+> - |+-\pm\pm> - |-+\pm\pm>]$$
(S9)

In organic semiconductors the lowest SE and TE levels are split ($\Delta_{ST}$) typically by ~0.7 eV due to the exchange interaction, J*. The TE is further split by ZFS typically of the order of ~10µeV. Therefore, to account for the splitting here we introduce anisotropic exchange interactions (AXI) within the four spin half system, which are large compared to D and E, but small compared to J* above. To keep the discussion relatively simple we describe here AXI between spins 1 and 2 ("first pair") and another AXI between spins 3 and 4 ("second pair"):

$$H_{x1} = E_{T1} - J_1(\vec{S}_1 \cdot \vec{S}_2 - 1/4) + \vec{S}_1 \cdot \tilde{V}_1 \cdot \vec{S}_2$$
$$H_{x2} = E_{T2} - J_2(\vec{S}_3 \cdot \vec{S}_4 - 1/4) + \vec{S}_3 \cdot \tilde{V}_2 \cdot \vec{S}_4$$
(S10)

where $J_i$ (<<J*) are the isotropic exchange and $V_i$ are 3x3 symmetric traceless tensors describing the anisotropy. For isotropic exchange, $\tilde{V}_i = 0$, each of the two equations in Eq. (S10) yields the triplet level of each of $H_{xi}$ at $E_{Ti}$ while the singlet level is at $E_{Ti}+J_i$.

Since the interactions are anisotropic, then the relative orientation of the two pairs affects the energy levels, wave functions, and spin configuration. Furthermore, the orientation of the external magnetic field, **B**, with respect to the pairs is also critical. In the laboratory frame of reference with B||z, the z-axis of each pair makes (θ,φ) spherical angles with B; the tensor $\tilde{V}$ (for each pair) has 5 components, and is given by an expression identical to Eq. (S6) except that *D* and *E* there should be replaced by 2*D* and 2*E*, respectively. The anisotropic part of the exchange $A = \vec{S}_1 \cdot \tilde{V} \cdot \vec{S}_2$ is thus composed of 5 additive terms:

$$A_0 = V_0[2S_{1z}S_{2z} - \frac{1}{2}(S_{1+}S_{2-} + S_{1-}S_{2+})]$$
$$A_{\pm 1} = V_{\pm 1}(S_{1\mp}S_{2z} + S_{1z}S_{2\mp})$$
$$A_{\pm 2} = V_{\pm 2}S_{\mp 1}S_{\mp 2}$$
(S11)

The full Hamiltonian of the four spin half system in a magnetic field B∥z is:

$$H_4 = \sum_{i=1}^{4} g_i \mu_B B S_{iz} + H_{x1}(\theta_1, \varphi_1) + H_{x2}(\theta_2, \varphi_2).$$
(S12)

When t-MPA of TT and SE are measured, the two spin pairs experience negative exchange and large J (>0 in Eqs. (S10)) and finite, but small compared to J* above, and $D$ and $E$: $|J|\gg D>E$. In this case we obtain three groups of energy levels: at $\sim E_{T1}+E_{T2}$ (TT=$S_1$+$T_1$+Q; 9 levels), at $\sim E_{T1}+ E_{T2} +J_i$ ($T_2$,$T_3$; 6 levels) and at ($E_{T1}+ E_{T2} +J_1+J_2$) ($S_2$, single level). The lowest energy group (the TT) is identified with CME. An enlarged view of the lowest group is shown in Fig. 4b, where we can see the ZFS and Zeeman split energy levels.

It is important to realize that the 9 CME levels at $E_{CME}=E_{T1} +E_{T2} \sim 2E_T$ is at energy twice the energy of the TE (of a single spin pair). In polymers with $E_T\sim E_S/2$ it is the TT group at energy $2*E_T\sim E_S$ which is near resonance with the measured SE and thus interact strongly producing the novel CME species.

Using ad-hoc ZFS parameters, $D$ and $E$, we calculated the 16 energy levels and wave functions of $H_4$ (Eq. (S12)) in $B$ applied in a general direction. We then calculated the t-MPA$_{S1}(B)$ and t-MPA$_{T1}(B)$, for $S_1$ and $T_1$ (Eqs. (S8, S9) above), respectively, using Eqs. (S2-S4) and averaging over all field directions. The results for two parallel and identical pairs of spin-half species (i.e., $H_{x1}$ and $H_{x2}$ are identical) are shown in Fig. 4b where it is seen that the $S_1$ and $T_1$ t-MPA(B) responses have opposite polarity but similar magnitude as observed in the t-MPA experiment.

It is important to note that the singlet and triplet contents of each level calculated either by the four spin ½ model or by the usual TT pair (a-la Merrifield) are different from each other; compare Figs. S1 and 4b. Thus, although the two approaches give rise to a TT-pair like states, these states are different in their wave functions. As a result, the t-MPA($B$) responses calculated using the 'four spin ½ model' are different from those calculated using the Merrifield TT Hamiltonian[33] (Eq. (S7)); see Figs. 4a and 4b.

**Additional References**

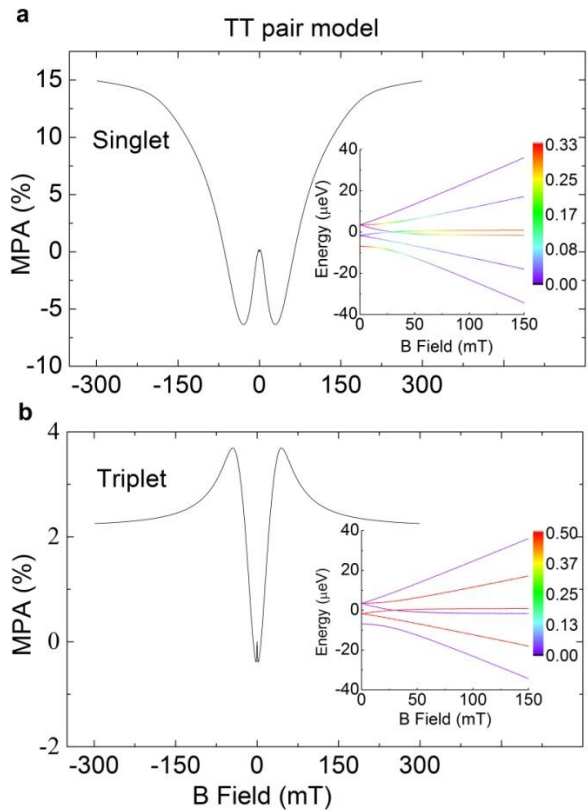

**Figure S.1| Calculated t-MPA(*B*) response using the Merrifield TT-pair model[33].** Calculated t-MPA(*B*) response at *t*=200 ps of the singlet (a) and triplet (b) components of the nine TT spin sublevels based on the same parameters as for the 'four spin ½ model' given in the text (see Fig. 4b). The insets show the energy levels, and singlet and triplet weights of the nine spin sublevels behind the MPA(*B*) response shown in (a) and (b).